\documentclass[fleqn,twoside]{article}
\usepackage{espcrc2}
\usepackage{amssymb}

\usepackage{graphicx}
\usepackage[figuresright]{rotating}

\newcommand{\mev}{\,{\rm MeV}}
\newcommand{\gev}{\,{\rm GeV}}
\newcommand{\chpt}{$\chi$PT}
\newcommand{\mpi}{m_\pi}

\title{
\vspace{-2.2cm}
\hfill \rm \null \hfill
\hbox{\normalsize ADP-03-129/T565} \\
\vspace{1.65cm}
Chiral Structure in Baryon Magnetic Moments}

\author{R.~D.~Young,
        D.~B.~Leinweber and 
        A.~W.~Thomas
\address[CSSM]{Centre for the Subatomic Structure
    of Matter and Department of Physics,\\
    University of Adelaide, Adelaide SA 5005 Australia}}
\begin{document}

\begin{abstract}
  We highlight recent advances in finite-range regularised chiral
  effective field theory. Application to baryon magnetic moments show
  an unambiguous signal of quenched chiral physics in FLIC fermion
  simulations.
\end{abstract}

\maketitle

\section{INTRODUCTION}
The issue of chiral extrapolation has been highlighted as an important
problem facing modern lattice QCD simulations
\cite{Bernard:2002yk,Thomas:2002sj}, if one is to make any
comparison with experiment. Extrapolations in quark mass are
nontrivial because of nonanalytic contributions which arise as a
consequence of dynamically broken chiral symmetry in the QCD vacuum
\cite{Detmold:2001hq}.

At typical lattice quark masses there is little
evidence of nonanalytic variation in quark mass. Recently, the
Kentucky group has reported observation of chiral nonanalytic
behaviour in quenched simulations with overlap fermions by reaching
pion masses as low as $180\mev$ \cite{Dong:2003im}.

The lack of rapid, nonanalytic curvature at moderate quark masses is
well understood through the use of a finite-range regulator (FRR) in
chiral effective field theory \cite{Young:2002ib,Donoghue:1998bs}.
Finite-range regularisation provides a resummation of the chiral
expansion which features the suppression of chiral loop effects with
increasing quark mass \cite{Leinweber:1999ig}, in accord with nature
where the finite size of the Goldstone boson source plays a role.

We first highlight the features of chiral extrapolation with the use
of FRR chiral perturbation theory (\chpt) in the case of the nucleon
mass in dynamical QCD. We discuss the quenched chiral extrapolation of
the magnetic moments of both nucleon and Delta baryons.

\section{FRR CHIRAL EFFECTIVE THEORY}
Chiral perturbation theory is a low-energy effective field theory
(EFT) for QCD. This EFT provides a systematic expansion of QCD about
the chiral limit. The key example here is that one can generate a
quark mass expansion for the behaviour of the nucleon mass. It is
therefore essential, for chiral extrapolations from moderate quark
masses, to incorporate the rigorous properties of QCD near the chiral
limit.

In the naive application of \chpt\ one obtains a truncated Taylor
series of the form
\begin{eqnarray}
m_N &=& c_0 + c_2 \mpi^2 + c_3 \mpi^3 + c_4 \mpi^4 \nonumber\\
    & & + c_{4L} \mpi^4 \log \mpi + \ldots .
\label{eq:exp}
\end{eqnarray}
The terms nonanalytic in the quark mass ($m_q\propto \mpi^2$) are
those arising strictly from pion loop integrals. They have
coefficients which are model-independent constraints of QCD. The
remaining analytic terms are determined by matching to
non-perturbative QCD. In principle, it is the goal of lattice QCD to
be able to constrain all coefficients --- including the reproduction
of the {\em correct} nonanalytic contributions. Until now, lattice
simulations have not observed the rapid nonanalytic variation and one
must fix the coefficients of these terms to their QCD values.  For
modern calculations, determination of the analytic terms enables
extrapolation to the physical pion mass.

The application of the formal expansion in Eq.~(\ref{eq:exp}) has
questionable applicability in the regime of modern QCD simulations,
$0.5\lesssim \mpi \lesssim 1.0\gev$, because the coefficients of
higher order terms are very large and oscillate in sign
\cite{Young:2002ib,Leinweber:2003ux}. This lack of convergence can be
overcome by using a FRR in chiral EFT. Chiral loop integrals are
evaluated with a finite cutoff in momentum space. This procedure
incorporates two additional {\em light} energy scales in the expansion
--- the $N$--$\Delta$ mass splitting and the finite size of the source
of the pion field. By incorporating both these scales and resumming
the chiral expansion one then has an enhanced range of applicability
of the effective field theory \cite{Young:2002ib}. This resummation
improves the convergence of the expansion to the point where the
chiral extrapolation of modern lattice simulations can now be
performed reliably \cite{Leinweber:2003dg}.

\begin{figure}[t]
\includegraphics[width=\columnwidth]{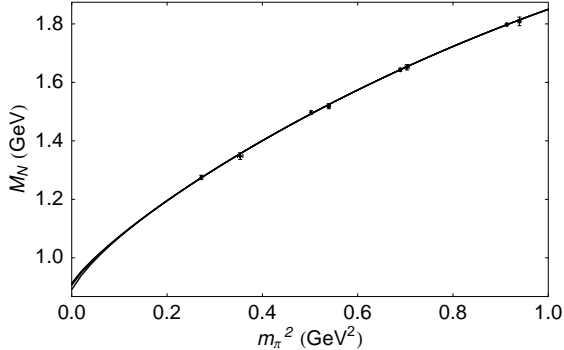}
\caption{Extrapolation of nucleon mass lattice data using four
  different finite-range regulators.}
\label{fig:mNext}
\end{figure}

By investigating a number of functional forms for the momentum-space
cutoff of loop integrals in chiral EFT it is found that the
extrapolated results show negligible sensitivity to the chosen form of
regulator. Results of the extrapolation of CP-PACS lattice data
\cite{AliKhan:2001tx} for four different regulators --- sharp,
monopole, dipole and gaussian --- are shown in Figure~\ref{fig:mNext}
\cite{Leinweber:2003dg}.

In summary, the resummation of the chiral series by the use of a FRR
provides a remarkably robust chiral extrapolation for lattice
simulations performed at moderate quark masses.

\section{MAGNETIC MOMENTS}
The study of magnetic moments on the lattice provides an enhanced
opportunity to directly study chiral nonanalytic physics. In QCD the
leading nonanalytic contribution to the nucleon magnetic moment is
linear in $\mpi$ --- compare $\mpi^3$ for the mass expansion,
Eq.~(\ref{eq:exp}). This contribution is large, nearly one third of
the physical neutron's magnetic moment. Such is the magnitude of the
pion loop effects that one should expect dramatic results in lattice
calculations as the light quark mass regime is probed.

Lattice simulations of the electromagnetic structure of baryons have
so far been restricted to the quenched model of QCD
\cite{Leinweber:1990dv,Gockeler:2003ay}. Because of the absence of
$q\bar{q}$ pair-creation, pion loop effects are typically suppressed
in the quenched approximation
\cite{Savage:2001dy,Leinweber:2001jc,Leinweber:2002qb}. The inclusion
of the flavour singlet $\eta'$ does offer the opportunity for
detection of enhanced chiral behaviour in quenched simulations. As a
result of double-hairpin loops the leading nonanalytic contribution is
$\log\mpi$ --- i.e. magnetic moments diverge in the chiral limit
\cite{Savage:2001dy}.

The Delta baryon offers a unique opportunity to study chiral loop
effects in quenched QCD. The $\Delta\to N\pi$ loop integral appears
with a negative sign, hence chiral curvature has the opposite effect
to that of QCD \cite{Leinweber:2003ux}. The enhancement of the
$N$--$\Delta$ mass splitting at light quarks in the quenched
approximation has been observed \cite{Young:2002cj,Leinweber:2002bw}.

The opposite sign on the $\Delta\to N\pi$ loop means that such a
vertex correction to the Delta baryon magnetic moment will also have a
sign which is opposite to that of QCD. Finite volume effects cannot
change this sign difference. The pion loop contributions to the
magnetic moments are $p$-wave and hence the main effect of the finite
volume is to create a gap in momentum space between $0$ and $2\pi/L$.
Because the loop
integrals are positive definite for both the proton and Delta when
$\mpi+m_N>m_\Delta$, the finite volume can only suppress the magnitude of
loop effects but cannot change their sign. Studying the nucleon and
Delta magnetic moments together therefore provides a clear signal of
quenched chiral physics independent of volume effects.

We have calculated the quenched chiral corrections to the Delta baryon
magnetic moment using the diagrammatic method of Leinweber
\cite{Leinweber:2001jc,Leinweber:2002qb}. The flavour symmetry of the
decuplet baryon interpolating fields means that the magnetic moments
are simply proportional to the charge in the quenched isospin symmetric limit.
Consideration of the $\Delta^{++}$ is therefore sufficient to
determine the quark mass dependence of the whole isobar.

The photon vertex correction from the double hairpin $\eta'$ dressing
$\Delta^{++}\to\Delta^{++}\eta'$ provides a logarithmic divergence.
The correction from a photon coupling to the meson loop trivially
vanishes in the quenched theory. In the case of the $\Delta^{++}$ a
pion loop is always a neutral $u\bar{u}$ pair and cannot carry a
charge current. It is therefore necessary to go to the next order
where the photon couples to the intermediate baryon state with a meson
loop {\it in the air}. The decuplet diagram,
$\Delta^{++}\to\Delta^{++}\pi^0$, is actually increased from the
analogous contribution in QCD by a factor $4/3$.

It is the decay channel to an intermediate octet state which gives
rise to the most interesting behaviour in the quenched theory. There
is a vertex correction from an intermediate $uuu$-octet state --- a
{\it double-charge ``proton''}, $p^{++}$ \cite{Leinweber:2003ux}. As
discussed by Cloet {\it et al.}, the opening of the decay channel at
$\mpi = m_\Delta-m_N$ gives
rise to dramatic curvature at pion masses near threshold
\cite{Cloet:2003jm}.

\begin{figure}[t]
\includegraphics[width=\columnwidth]{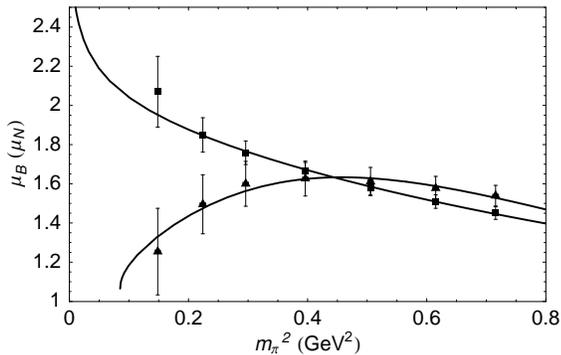}
\caption{Preliminary lattice results for proton ($\blacksquare$) and
  $\Delta^+$ ($\blacktriangle$) magnetic moments. Curves represent
  best fit to data using FRR Q\chpt.}
\label{fig:mag}
\end{figure}

We show preliminary results of FLIC fermion simulations
\cite{Zanotti:2001yb} of baryon magnetic moments \cite{JZlat03}. In
particular, we plot together the $p$ and $\Delta^+$ magnetic
moments in Figure~\ref{fig:mag}.  We observe that at moderate quark
masses the expected heavy-quark theory result is observed, with the
$\Delta^+$ lying slighty above the proton.
However, in the light-quark regime chiral physics becomes
apparent with the clear separation of the two signals. The downward
curvature in the $\Delta^+$ is dominated by the coupling to an
intermediate octet baryon.

The opposite behaviour of the proton and Delta allows one to
unambiguously identify quenched chiral physics at pion masses
$\lesssim 500\mev$.

This work was supported by the Australian Research Council and the
University of Adelaide.

\end{document}